\documentclass[twocolumn,trackchanges]{aastex631}

\received{\today}
\revised{--}
\accepted{--}

\usepackage[utf8]{inputenc}
\usepackage{graphicx}
\usepackage{bm}
\usepackage{color}      
\usepackage{units}
\usepackage{amsmath}
\usepackage{epsfig}
\usepackage{amssymb}
\usepackage{mathrsfs}
\usepackage{xcolor}
\colorlet{RED}{red}

\allowdisplaybreaks
\def\be{\begin{align}}
\def\ee{\end{align}}
\def\bea{\begin{eqnarray}}
\def\eea{\end{eqnarray}}
\def\ban{\begin{eqnarray*}}
\def\ean{\end{eqnarray*}}
\def\bd{\begin{displaymath}}
\def\ed{\end{displaymath}}

\def\bc{\begin{center}}
\def\ec{\end{center}}

\def\ba{\begin{array}}
\def\ea{\end{array}}

\shorttitle{Formation of the hammerhead ion populations observed by PSP}
\shortauthors{Shaaban et al.}

\begin{document}

\title{Decoding the formation of hammerhead ion populations observed by Parker Solar Probe}

\author[0000-0002-0786-7307]{Shaaban M. Shaaban}
\affiliation{Department of Physics and Materials Sciences, College of Arts and Sciences, Qatar University, 2713 Doha, Qatar.}
\author[0000-0002-8508-5466]{M. Lazar}
\affiliation{Centre for Mathematical Plasma Astrophysics, Dept. of Mathematics, KU Leuven, Celestijnenlaan 200B, 3001 Leuven Belgium.}
\affiliation{Institute for Theoretical Physics IV, Faculty for Physics and Astronomy, Ruhr-University Bochum, D-44780 Bochum, Germany.}
\author[0000-0003-3223-1498]{R. A. L\'opez}
\affiliation{Research Center in the intersection of Plasma Physics, Matter, and Complexity ($P^2 mc$), Comisi\'on Chilena de Energ\'{\i}a Nuclear, Casilla 188-D, Santiago, Chile}
\affiliation{Departamento de Ciencias F\'{\i}sicas, Facultad de Ciencias Exactas, Universidad Andres Bello, Sazi\'e 2212, Santiago 8370136, Chile}

\author[0000-0001-8134-3790]{P. H. Yoon}
\affiliation{Institute for Physical Science and Technology, University of Maryland, College Park, MD 20742-2431, USA}
\author[0000-0002-1743-0651]{S. Poedts}
\affiliation{Centre for Mathematical Plasma Astrophysics, Dept. of Mathematics, KU Leuven, Celestijnenlaan 200B, 3001 Leuven Belgium.}
\affiliation{Institute of Physics, University of Maria Curie-Sk{\l}odowska, Pl.\ M.\ Curie-Sk{\l}odowska 5, 20-031 Lublin, Poland.}

\begin{abstract}
In situ observations by the Parker Solar Probe (PSP) have revealed new properties of the proton velocity distributions, including hammerhead features that suggest non-isotropic broadening of the beams. 
The present work proposes a very plausible explanation for the formation of these populations through the action of a proton firehose-like instability triggered by the proton beam.
The quasi-linear (QL) theory proposed here shows that the resulting right-hand (RH) waves have two consequences on the protons: (i) reduce the relative drift between the beam and the core, but above all, (ii) induce a strong perpendicular temperature anisotropy, specific to the observed hammerhead ion strahl.
Moreover, the long-run QL results suggest that these hammerhead distributions are rather transitory states, still subject to relaxation mechanisms, of which instabilities like the one discussed here are very likely involved.

\end{abstract}


\section{Introduction} \label{sec:intro}

New in situ observations by the Parker Solar Probe (PSP) in the young solar wind confirm an extended presence of the core-beam tandem in the observed proton velocity distributions \citep{Klein_2021, Verniero-etal-2022}.
The theoretical and numerical modeling of these populations thus becomes even more motivating, especially for understanding the origin of proton beams, but also their implications in the solar wind plasma dynamics \citep{Klein_2021, Verniero-etal-2022, Ofman_2022}.
Existing scenarios, not necessarily in consensus, predict that proton beams can be injected by magnetic reconnection events at the coronal base but can also result from the trapping and acceleration of protons by resonant waves, e.g., cyclotron or even kinetic Alfven waves \citep{Marsch-2006, Araneda-etal-2008, Pierrard-Voitenko-2010}.
An immediate consequence of proton beams is self-generated wave instabilities, which are expected to convert bulk kinetic energy to smaller scales because the resulting wave fluctuations can subsequently dissipate and heat the solar wind plasma \citep{Marsch-2006, Bale-etal-2019, Bowen_2020}.
Favorable evidence emerges from the PSP data, both regarding the source of the proton beams and their regularization by self-induced instabilities \citep{Bowen_2020, Verniero-etal-2022, Phan-etal-2022}.
Moreover, observations of proton beams simultaneous with ion-scale wave fluctuations \citep{Klein_2021, Verniero-2020} complement data from, e.g., Helios, Wind and STEREO, and facilitate solving a long debate on the dominance of right-handed (RH) magnetosonic (MS) modes over left-handed (LH) ion-cyclotron (IC) waves \citep{Marsch-1982, Daughton-1998, Gary-2016, Woodham_2019}.
Depending on their properties, the proton beams can destabilize either RH-MS modes or LH-IC instability, of which the generation of RH wave fluctuations seems more complex and is still an open question \citep{Verniero-2020}.

In this letter, we show that the proton-beam firehose-like instabilities of the RH transverse waves can determine a broad relaxation of the beams, resembling the hammerhead distributions reported by PSP \citep{Verniero-etal-2022}.
Section 2 explains the quasi-linear (QL) formalism of the transverse instabilities induced by the drifting bi-Maxwellian proton population. 
The parameterization used for the proton populations is realistic and consistent with observations.
In section 3, we discuss several relevant results from our parametric analysis and provide a semi-quantitative comparison with the observations.
Growing waves partially convert the proton beam's free (kinetic) energy into the heating of this component, mainly in the perpendicular direction.
The properties of RH waves and the resulting temperature anisotropy, $A_b = T_{b \perp}/T_{b \parallel} > 1$ ($\parallel, \perp$ are gyrotropic directions with respect to the background magnetic field), also conform to the observations \citep{Verniero-etal-2022}.
The last section summarizes the results and draws the main conclusions of this Letter.


\section{QL approach of proton beam instability} \label{sec3}

In the proton (subscript $p$) velocity distribution  
\begin{equation}\label{eq1}
f_p\left(v_\parallel,v_\perp \right)=\dfrac{n_c}{n_p}~f_c\left(v_\parallel,v_\perp 
\right)+ \dfrac{n_b}{n_p} ~f_b\left(v_\parallel,v_\perp \right).  
\end{equation}
both the core (subscript $c$) and beam (subscript $b$) components are assumed well described by drifting bi-Maxwellian distributions
\begin{align} \label{eq2}
f_{j}\left(v_{\parallel },v_{\perp }\right) =&\frac{\pi
^{-3/2}}{\alpha_{\perp j}^{2} \alpha_{\parallel j}}\exp \left[-\frac{v_{\perp}^{2}}{\alpha_{\perp a}^{2}}-\frac{\left(v_{\parallel }-v_j\right)^{2}} {\alpha_{\parallel a}^{2}}\right].   
\end{align}
$n_c$ and $n_b$ are the core and beam number density, respectively, and 
$n_p=n_b+n_c$ is the total number density of protons. $\alpha_{j\parallel, \perp}(t)=[2k_B T_{j \perp, \parallel}(t)/m_p]^{1/2}$ are components of thermal velocities, while $v_j$ are drift velocities, which preserve a zero net current $n_c v_c = n_b v_b$.
According to our QL approach, these velocities vary in time ($t$) as a measure of energy exchange.
Electrons (subscript $e$) ensure a neutral plasma ($n_e = n_p$). They are non-drifting ($v_e=0$) and initially, isotropic and Maxwellian distributed.

For this plasma configuration, the (instantaneous) dispersion relation for the RH transverse modes propagating parallel to the magnetic field reads \citep{Shaaban-etal-2020}
\begin{align} \label{eq3}
\tilde{k}^2=&\delta_c\left[\Psi_c+\Gamma_{c+}^+ Z_{c}\left(\xi_{c}^{+}\right)\right]+\delta_b\left[\Psi_b+\Gamma_{b-}^+ Z_{b}\left(\xi_{b}^{+}\right)\right]\nonumber\\
&+~\mu \left[\Psi_e+\Gamma_e^- Z_{e}\left(\xi_{e}^{-}\right)\right],
\end{align} 
$\tilde{k}=ck/\omega_{p p}$ is the normalized wave-number, $c$ is the speed of light, $\omega_{pp}=(4\pi n_p e^2/m_p)^{1/2}$ is the proton plasma frequency, $\tilde{\omega}=\omega/\Omega_p$ is the normalized wave frequency, $\Omega_j=~e B_0/m_j c$ is the non-relativistic gyro-frequency of the plasma species $j$ (with elementary charge $e$), $\beta_{j \parallel}=8 \pi n_j k_B T_{j \parallel}/B_0^2$ are the parallel plasma beta parameters, $\Psi_j=A_j-1$, $A_j=T_{j \perp}/T_{j \parallel}\equiv\beta_{j \perp}/\beta_{j \parallel}$ is the temperature anisotropy, $\mu=m_p/m_e$ is the proton-electron mass ratio, $U_j=V_j/\sqrt{\delta_c}$, $V_b=v_b/v_{Ac}$, $v_{Ac}=B_0/\sqrt{4\pi n_c m_p}$ is the proton core Alfv{\'e}n speed, $Z_{j}\left(\xi_{j}^{\pm}\right)$
%
%
is the plasma dispersion function \citep{Fried1961} of arguments
\begin{align*}
\xi_{j \mp}^{+}=\frac{\tilde{\omega}+1\mp\tilde{k} U_j}{\tilde{k} \sqrt{\beta_{j \parallel}/\delta_j}},~~
\xi_{e}^{-}=\frac{\tilde{\omega}-\mu}{\tilde{k} \sqrt{\mu\beta_{e \parallel}}},
\end{align*}  
and
\begin{align*}
\Gamma_{j\mp}^+=&\frac{A_j(\tilde{\omega}+1\mp\tilde{k} U_j )-1}{\tilde{k} \sqrt{\beta_{j \parallel}/\delta_j}},~     \Gamma_{e}^-=\frac{A_e\left(\tilde{\omega}-\mu\right)+\mu}{\tilde{k} \sqrt{\mu \beta_{e \parallel}}}
\end{align*}
The QL evolution of the macroscopic parameters (moments of $f_j$), such as the plasma betas $\beta_{\perp, \parallel j}$ and drifting velocities $V_j$, is governed by the following equations
\begin{subequations}\label{eq5}
\begin{align}
\frac{d\beta_{j \perp}}{d\tau}=&-\delta_j\int\frac{d\tilde{k}}{\tilde{k}^2} W(\tilde{k})\left[\Lambda_j \tilde{\gamma}+ G^{\pm}_{j \perp}~\eta_{j \mp}^{\pm}\right],\label{eq9a}\\
\frac{d\beta_{j \parallel}}{d\tau}=&2 \delta_j\int\frac{d\tilde{k}}{\tilde{k}^2} W(\tilde{k}) \left[A_j~\tilde{\gamma}+G_{j \parallel}~\eta_j^{\pm}\right],\label{eq9b}\\
\frac{d V_b}{d\tau}=&\frac{\sqrt{\delta_c}}{2}\int\frac{d\tilde{k}}{\tilde{k}} W(\tilde{k}) 
\text{ Im}~\eta_{b-}^{+}/\left(\tilde{k}\sqrt{\beta_{b\parallel}/\delta_b}\right),\\
\frac{d V_c}{d\tau}=&\frac{-\sqrt{\delta_c}}{2}\int\frac{d\tilde{k}}{\tilde{k}} W(\tilde{k}) 
\text{ Im}~\eta_{c+}^{+}/\left(\tilde{k}\sqrt{\beta_{c \parallel}/\delta_c}\right),
\end{align}
\end{subequations}
in terms of $\tau=\Omega_pt$, the instantaneous growth rate $\tilde{\gamma}(t) =\gamma/\Omega_p$ derived from the linear dispersion relation \eqref{eq3}, the normalized wave (spectral) energy density $W(\tilde{k})=\delta B^2(\tilde{k})/B_0^2$, and
\begin{subequations}\label{eq6}
\begin{align}
\delta_j=n_j/&n_e, ~\Lambda_j=\left(2 A_j-1\right), ~\Lambda_e=\mu\left(2 A_e-1\right)\\
\eta^+_{j\mp}=&\left[A_j\left(\tilde{\omega}\mp\tilde{k} U_j\right)+\left(A_j-1\right)\right]Z_{j}\left(\xi_{j \mp}^{+} \right),\\
\eta_e^{-}=&\mu \left[A_e~\tilde{\omega}-\left(A_e-1\right)\mu\right]Z_{e}\left(\xi_e^-\right),\\
G_{j \perp}^+=&\text{Im} \frac{2i\tilde{\gamma}+ 1}{\tilde{k}\sqrt{\beta_{j \parallel}/\delta_j}}, ~~~~G_{e \perp}^-=\text{Im} \frac{2i\tilde{\gamma}-\mu}{\tilde{k}\sqrt{\mu\beta_{e \parallel}}},\\
G_{j \parallel\mp}^+=&\text{ Im} \frac{\tilde{\omega}+1\mp \tilde{k} U_j}{\tilde{k}\sqrt{\beta_{j \parallel}/\delta_j}},~~ G_{e \parallel}^{-}= \text{Im} \frac{\tilde{\omega}-\mu}{\tilde{k}\sqrt{\mu\beta_{e \parallel}}}.
\end{align}
\end{subequations}
The wave equation completes the QL equations
\begin{equation} \label{eq7a}
\frac{\partial~W_t(\tilde{k})}{\partial \tau}=2~\tilde{\gamma}(\tau)~ W_t(\tilde{k}).
\end{equation}
\begin{figure*}[t!]
    \centering
    \includegraphics[width=0.245\textwidth]{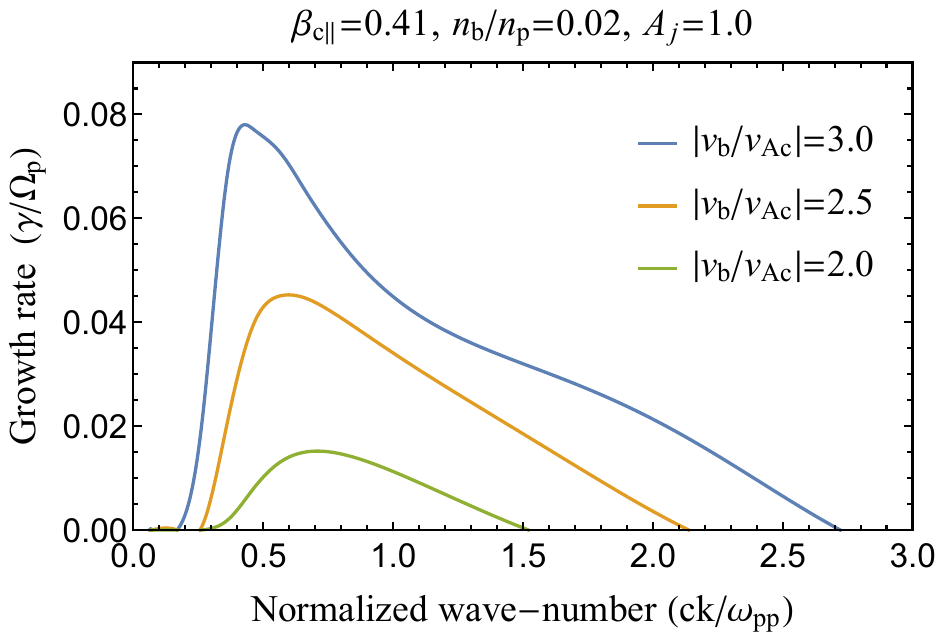}
    \includegraphics[width=0.245\textwidth]{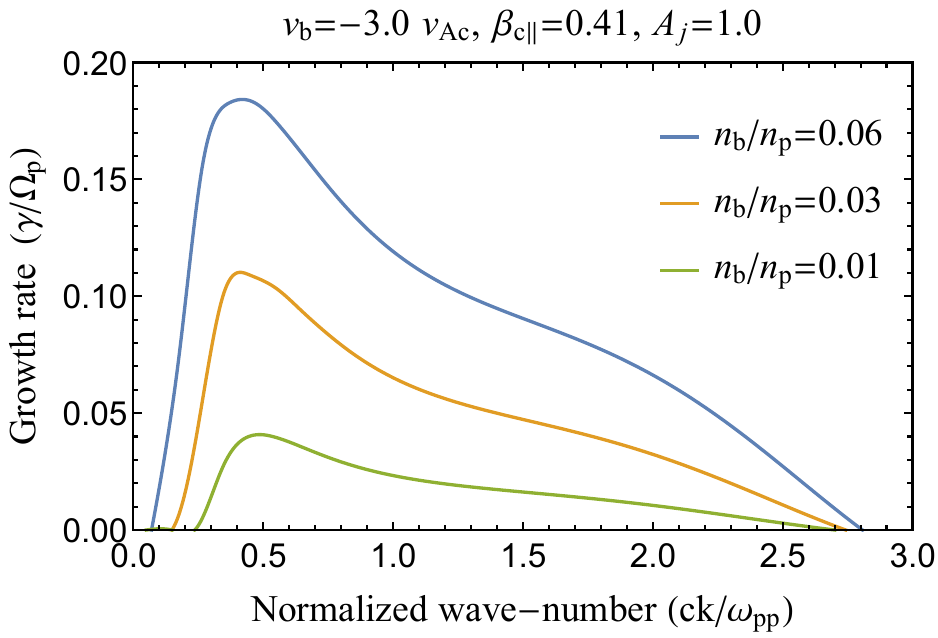}
    \includegraphics[width=0.245\textwidth]{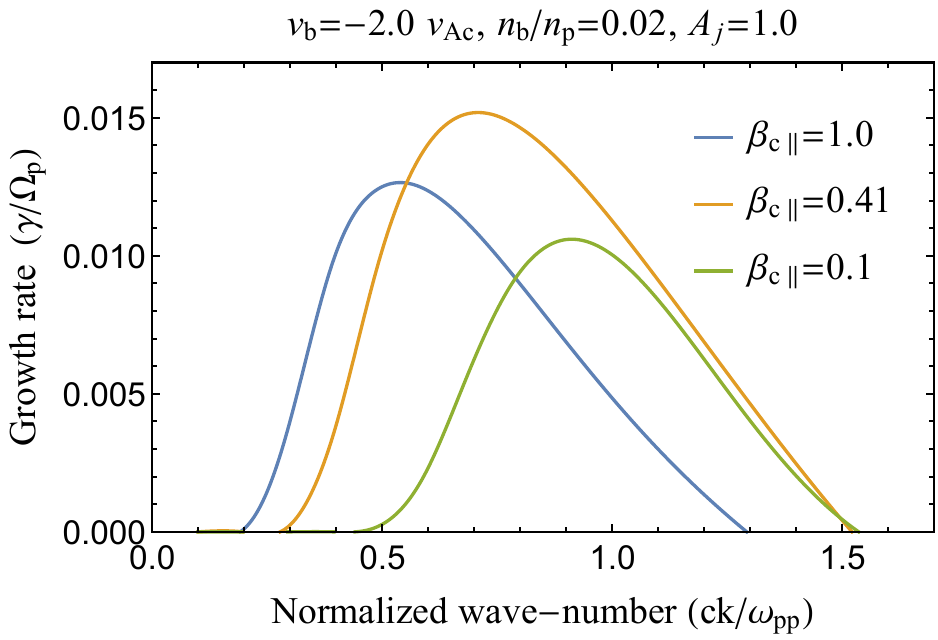}
    \includegraphics[width=0.245\textwidth]{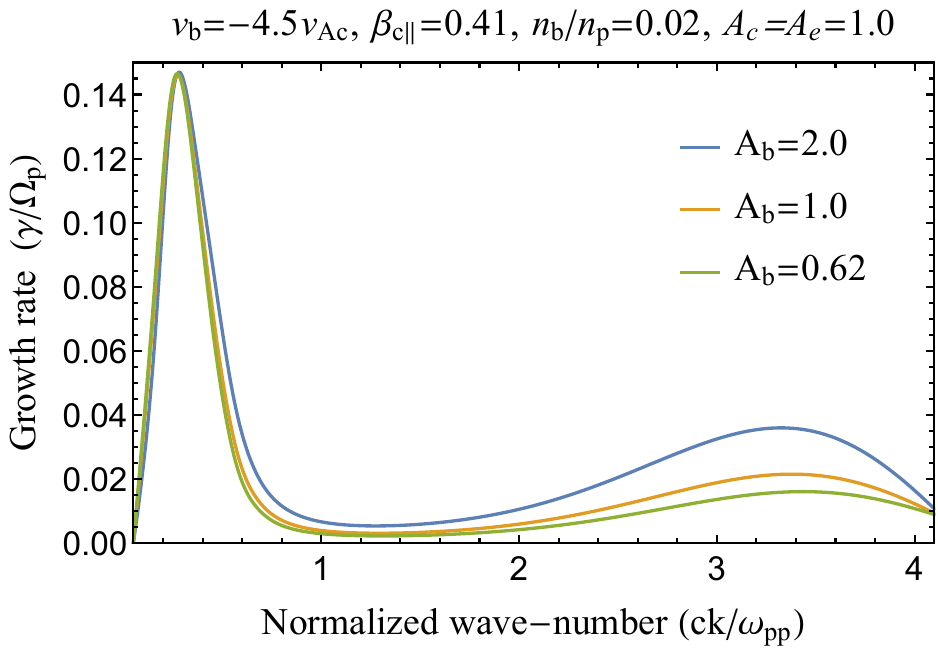}\\
    \includegraphics[width=0.245\textwidth]{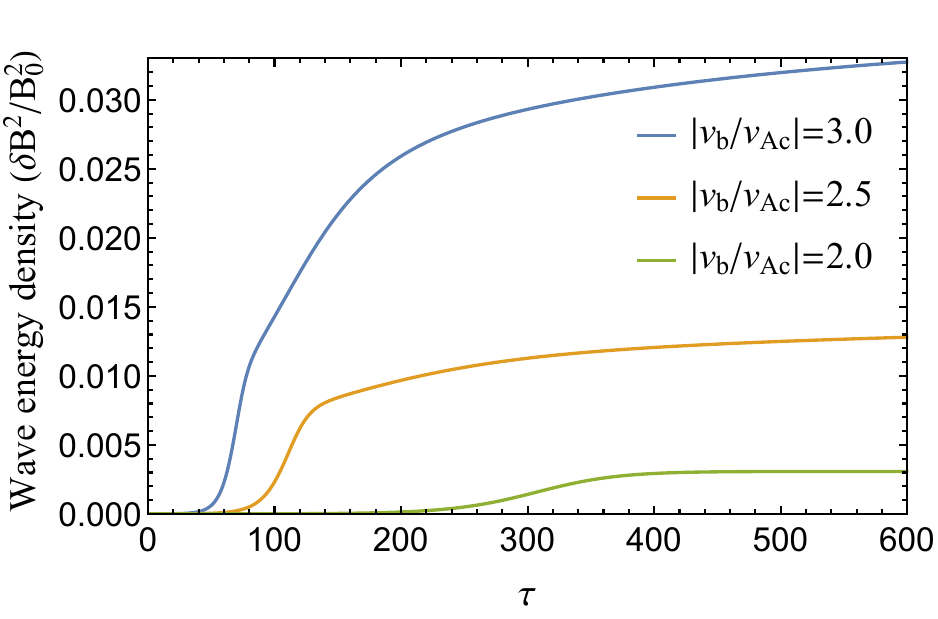}
    \includegraphics[width=0.245\textwidth]{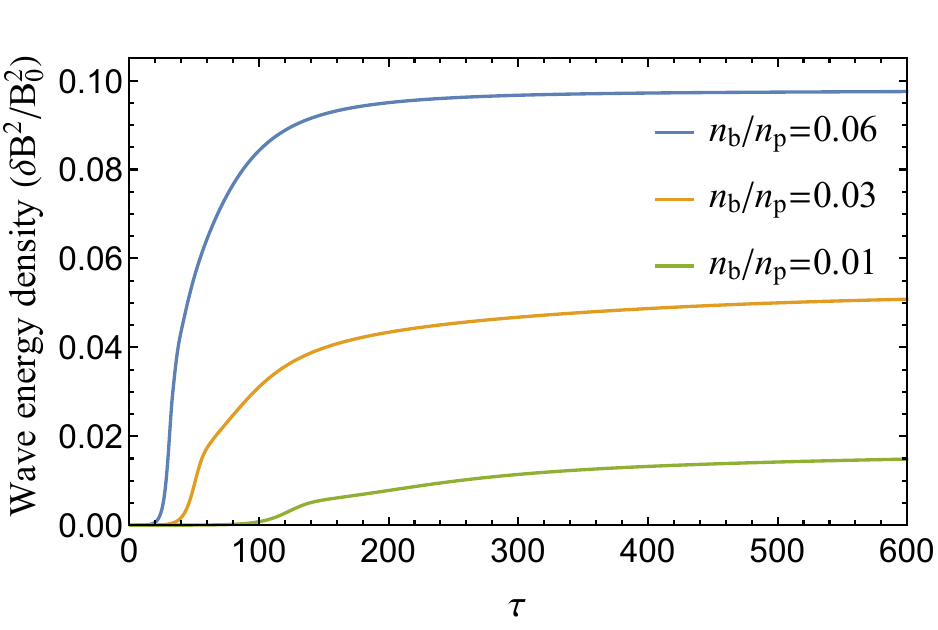}
    \includegraphics[width=0.245\textwidth]{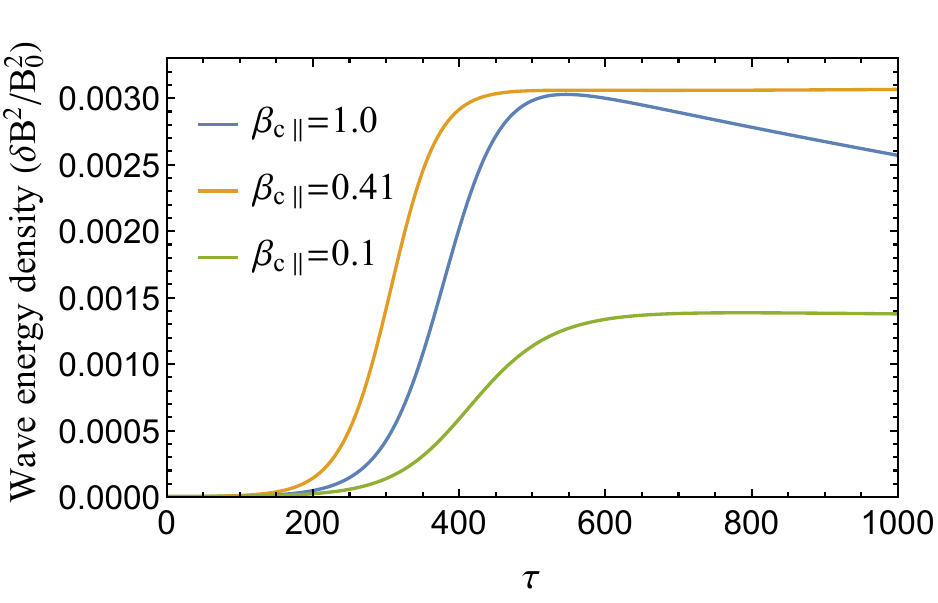}
     \includegraphics[width=0.245\textwidth]{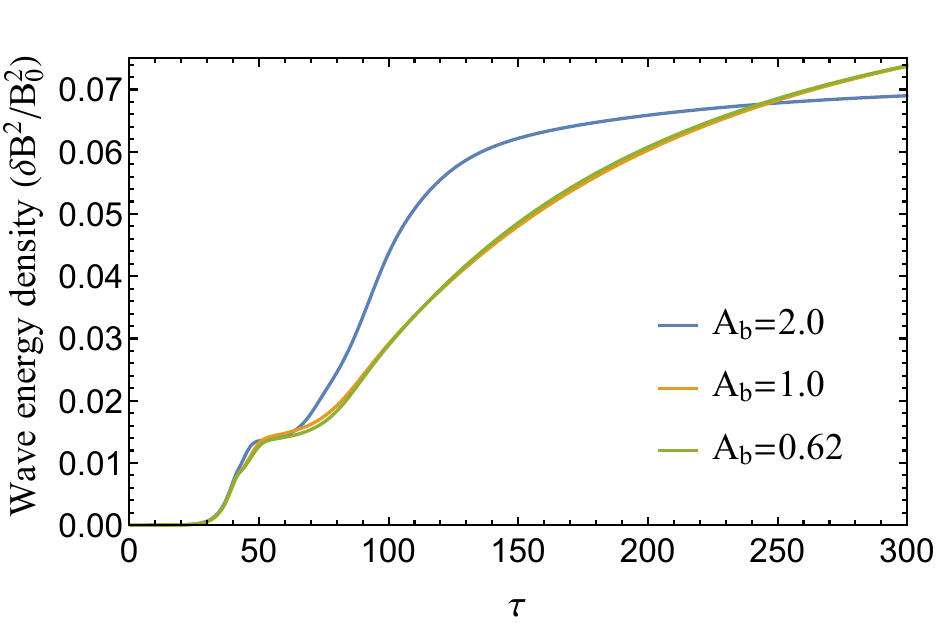}\\
    \includegraphics[width=0.245\textwidth]{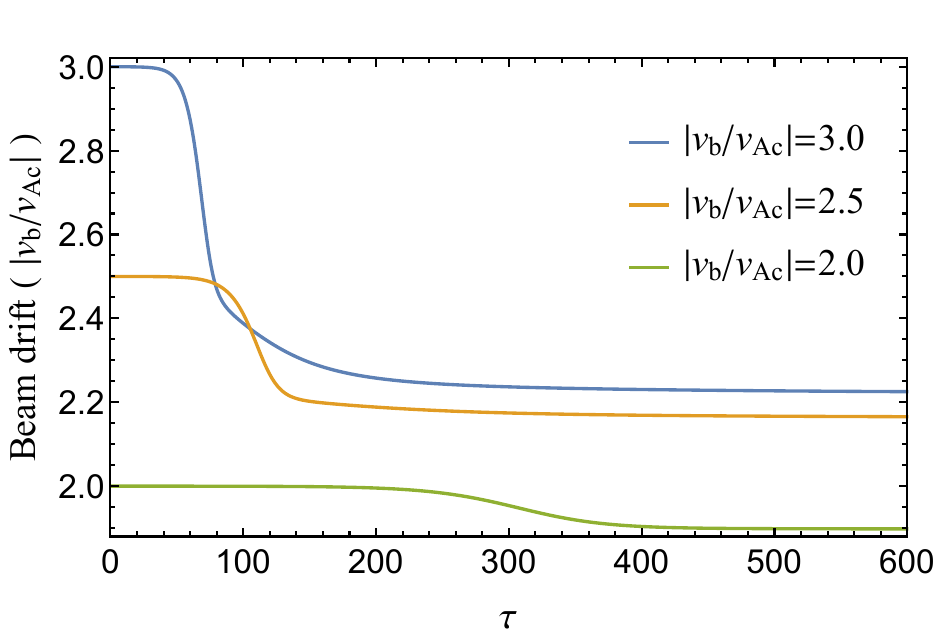}
    \includegraphics[width=0.245\textwidth]{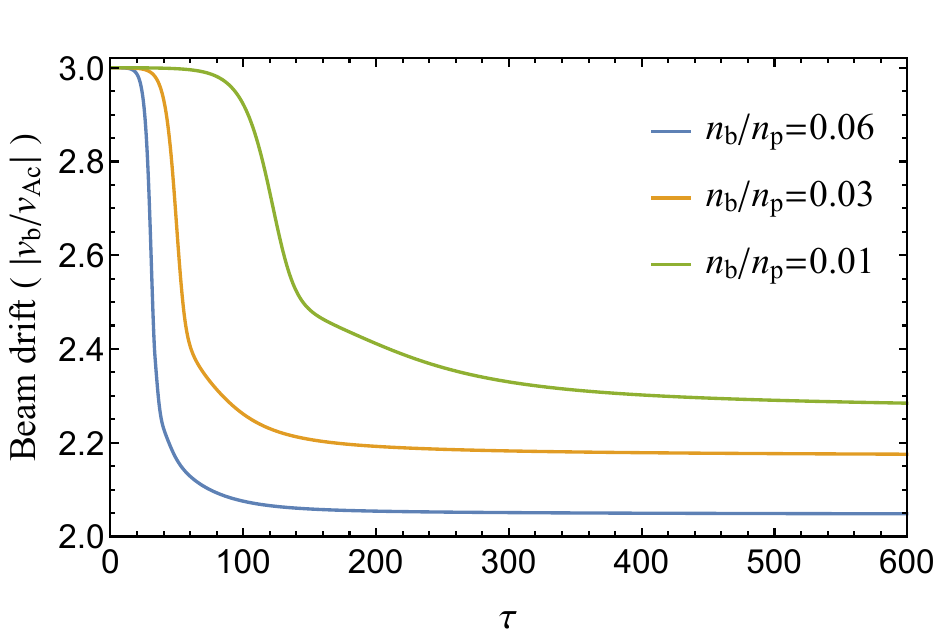}
    \includegraphics[width=0.245\textwidth]{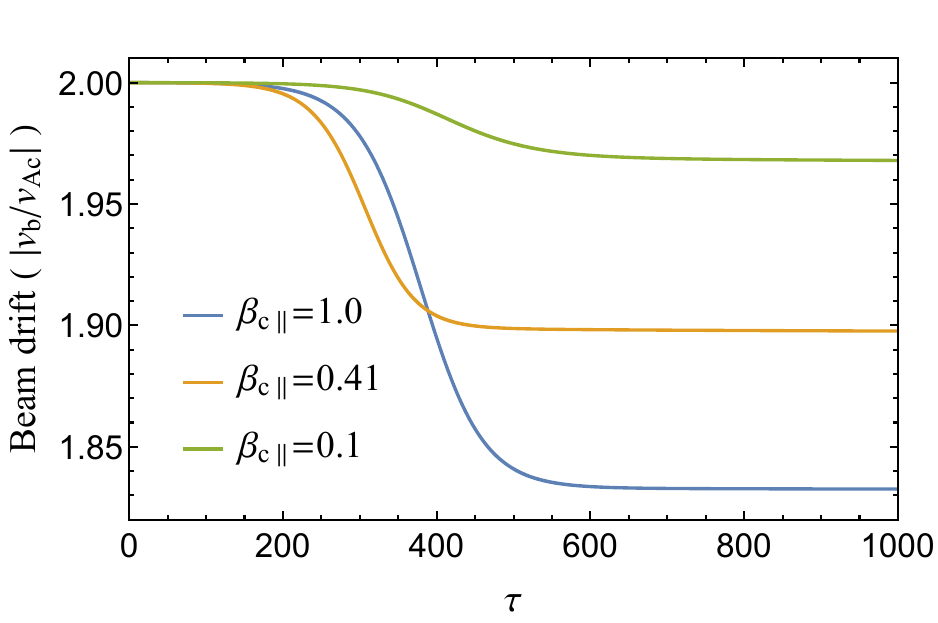}
    \includegraphics[width=0.245\textwidth]{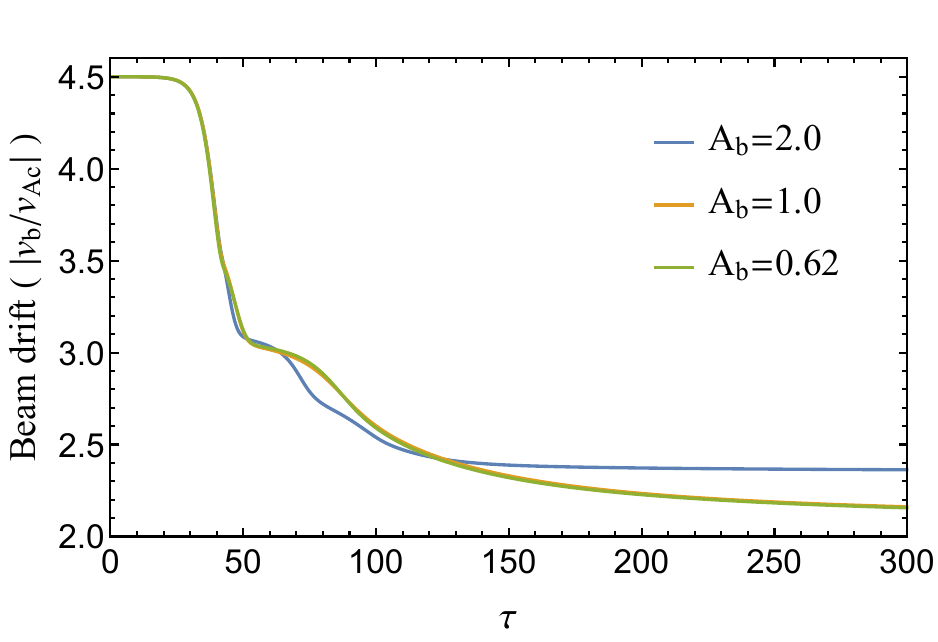}\\
    \includegraphics[width=0.245\textwidth]{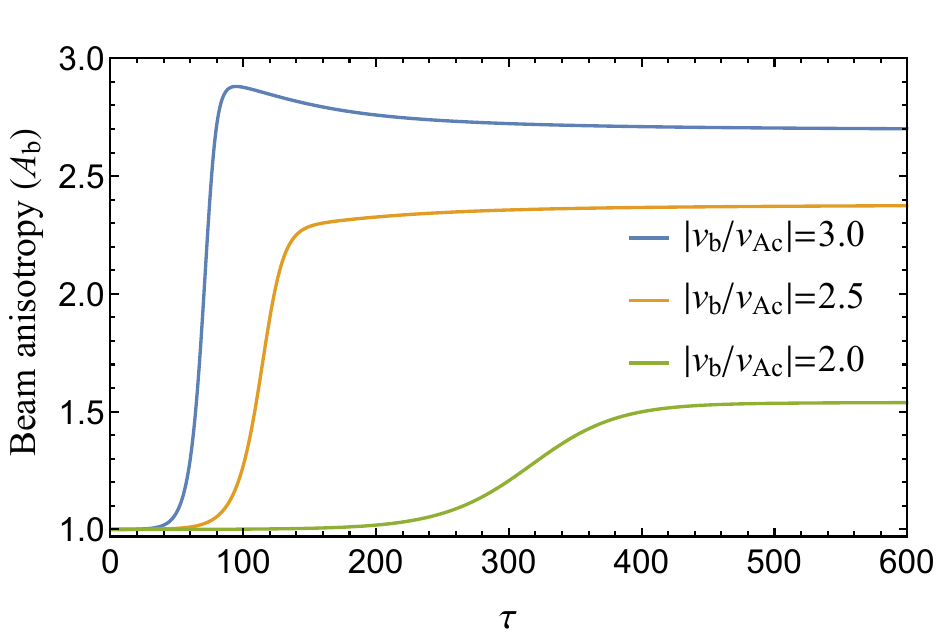}
    \includegraphics[width=0.245\textwidth]{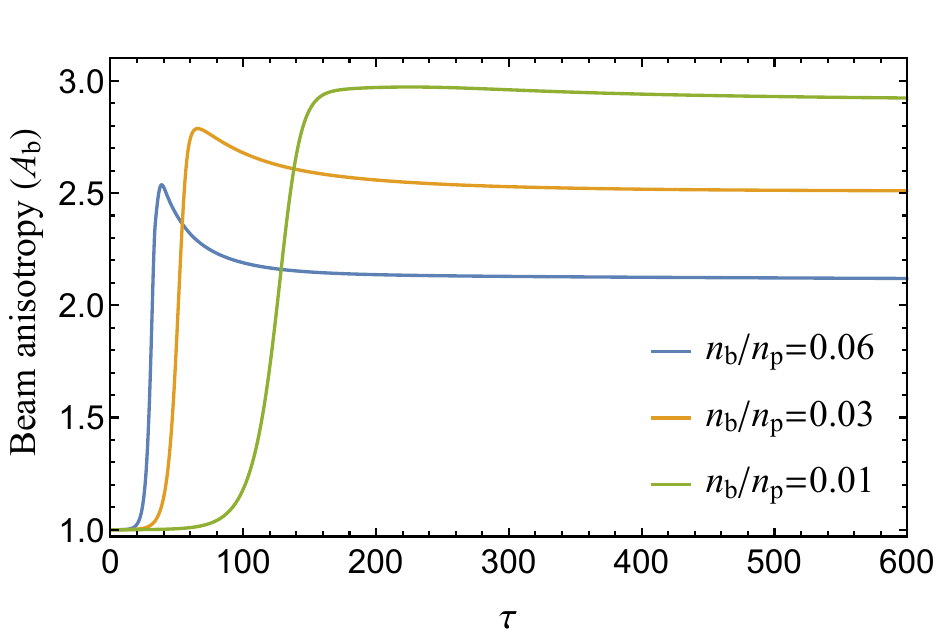}
    \includegraphics[width=0.245\textwidth]{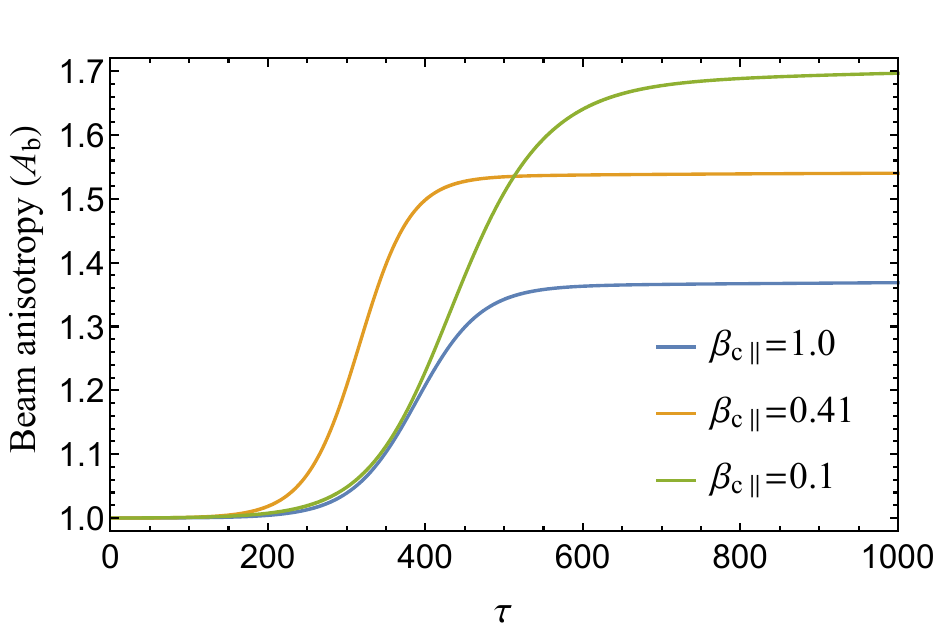}
    \includegraphics[width=0.244\textwidth]{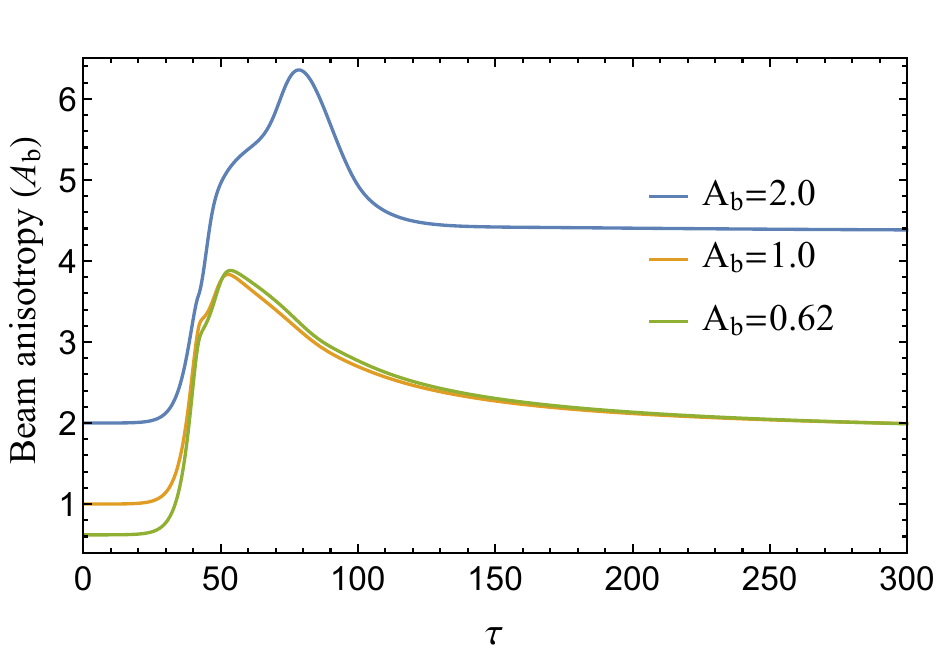}
    \caption{Growth rates ($\tilde{\gamma}$) of the  RHPB instability (top), and the corresponding time evolutions for the wave energy density (second upper row), beam velocity (third row), and the beam temperature anisotropy (bottom) for different initial conditions: $v_b/v_{Ac}=3.0, 2.5, 2.0$ (first-left column), $n_b/n_p=0.06, 0.03, 0.01$ (second column), $\beta_{c\parallel}=1.0, 0.41, 0.1$ (third column), and $A_b=2.0, 1.0, 0.62$ (right column).}
    \label{fig:1}
\end{figure*}
%

\section{Results supporting the formation of the hammerhead proton beam}\label{sec:res}
%
Here, we discuss the numerical results from linear and QL analysis for 15 runs with different initial parameters: 
\begin{align}
&\text{Runs 1, 2, 3: }|V_b(0)|=3.0, 2.5, 2.0,\nonumber\\
&\text{Runs 4, 5, 6: } |V_b(0)|=3.0,~n_b/n_p=0.06, 0.03, 0.01,\nonumber\\
&\text{Runs 7, 8, 9: }|V_b(0)|=2.0,~\beta_{c\parallel}(0)=1.0, 0.41, 0.1, \nonumber \\
&\text{Runs 10, 11, 12: }|V_b(0)|=4.5,~A_b(0)=2.0, 1.0, 0.62.\nonumber\\
&\text{Runs 13, 14, 15 in Figure \ref{fig:3}: } |V_b(0)|= 4.5.0, 4.0, 3.5.\nonumber
\end{align}
The other initial values used for the plasma parameters are $W(\tilde{k})=10^{-6}$, $A_j(0)=1$, $n_b/n_p=0.02$, $\beta_{c\parallel}(0)=0.41$, and $T_{b\parallel}/T_{c\parallel} = 2.465$, unless otherwise specified. 
These are estimations from the PSP observations in, e.g., \cite{Klein_2021, Verniero-etal-2022}.

\subsection{Linear analysis}
The dispersion relation \eqref{eq3} is solved exactly numerically to derive the unstable solutions of the RH proton-beam (RHPB) mode, whose growth rates ($\tilde{\gamma}$) are displayed in Figure~\ref{fig:1} (the first-row upper panels) as a function of the wave number ($\tilde{k}$). 
Growth rates are systematically enhanced, especially with increasing the beam drift velocity $|v_b/v_{Ac}|$ (first panel left) and beam density $n_b/n_p$ (second panel). 
Stimulation also consists of expanding the range of unstable mode wavenumbers.
The increase of beam temperature in the perpendicular direction has a similar but less significant effect (fourth panel). If the beam velocity is larger, the growth rate may combine both peaks of RH and LH modes \citep{Shaaban-etal-2020}.
The variation of growth rates is less monotonic, as they increase and then decrease with the plasma beta (third panel). 
This is specific to beam instabilities of electromagnetic or transverse modes \citep{Shaaban-etal-2018, Shaaban-etal-2020}. 
In this case, we deal with a proton beam firehose-like instability that develops between two threshold conditions, between the regime of electrostatic instabilities for a low beta parameter (or thermal spread) and the regime of ion-cyclotron instabilities for a high beta (or thermal spread).
The wave frequency of the unstable mode (not shown here) displays minor variations when the different plasma parameters are changed. 
Note also that the unstable solutions derived from the RH dispersion relation \eqref{eq3}, i.e., with $\xi_{b,c}^{+}$, have positive wave frequencies ($\omega_r/\Omega_p>0$) for $\tilde{k}>0$, confirming their RH polarization.

\begin{figure}[h!]
    \centering
    \includegraphics[width=0.36\textwidth]{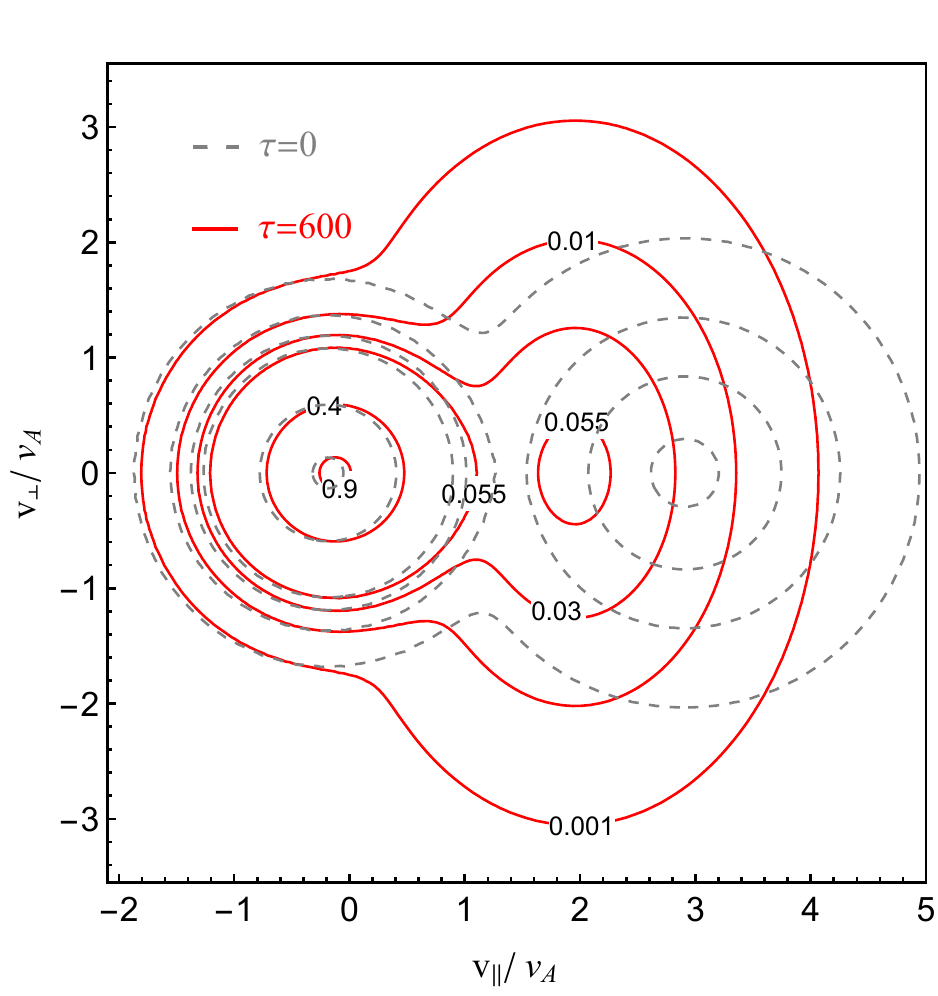}
    \caption{Contours of proton velocity distribution for the parameters in run~4, initially at $\tau=0$ (gray-dashed) and at the final running time $\tau=600$ (red-solid), when the distribution resembles the hammerhead shapes reported by PSP.}
    \label{fig:2}
\end{figure}
\subsection{QL analysis}
Beyond linear theory, we resolve the set of QL equations \eqref{eq5}--\eqref{eq7} for the initial parameters (i.e., at $\tau=0$) of each run indicated above. 
The results are also shown in Figure~\ref{fig:1}, first, the time evolution of the RHPB wave energy density $W_t(\tau)\equiv\delta B^2/B_0^2$ (second row upper). 
Then we describe the back reaction of the growing waves on the macroscopic plasma parameters, such as the beam velocity $|v_b(\tau)/v_{Ac}|$ (third row), and the induced beam temperature anisotropy $A_b(\tau)$ (bottom). 
Each of them corresponds to the initial growth rates displayed in the first upper panels, with the same color style.

%
Let's first discuss runs~1-3, aimed to show the variations with the beam velocity, and whose results are displayed in the left column of Figure~\ref{fig:1}, for $|v_b(0)/v_{Ac}|=$ 3.0 (blue), 2.5 (orange), and 2.0 (green).
An increase of the initial $|v_b(0)/v_{Ac}|$ can determine not only a faster initiation (blue line) but also higher levels of the RHPB fluctuations (second upper panel), confirming a stimulation of the instability predicted by linear theory (top panel).
The excitation of RHPB fluctuations regulates the free energy of the proton beam populations, reducing the beam velocity or relative drift (third panel), but also through the beam population's preferential cooling and heating mechanisms. 
The relaxation of the drift velocity is accompanied by perpendicular heating of the beam and a relatively large anisotropy $T_{b\perp}>T_{b\parallel}$ (bottom panel) reached at (or slightly before) the instability saturation. 
These effects are more prominent and faster for more energetic beams with a higher initial velocity.
For example, the beam initially having  $|v_b(0)/v_{Ac}|=3.0$ loses $\sim 25\%$  after relaxation, whereas those with $|v_b(0)/v_{Ac}|=2.5$ and 2.0 lose $\sim 14\%$ and $\sim 9\%$, respectively. 
Moreover, the initially isotropic protons of the beam component, with $A_b(0)=1.0$, experience a stronger perpendicular heating (blue line) if their initial $|v_b(0)/v_{Ac}|$ is higher, resulting in a significant induced temperature anisotropy at later stages, i.e., $A_b(\tau_{max}) \simeq 2.9$ for $|v_b(0)/v_{Ac}|=3.0$.

%
%
The second (left) column in Figure~\ref{fig:1} displays the results from runs 4, 5, and 6, corresponding to different initial number densities of the beam, $n_b/n_p=$ 0.06 (blue), 0.03 (orange), and 0.01 (green): 
The temporal profile of the wave energy density of RHPB fluctuations (second upper panel), and their back reactions on the beam drift velocity $|v_b(\tau)/v_{Ac}|$ (third panel) and the beam temperature anisotropy $A_b(\tau)$ (bottom).
Denser beams are more efficient in shortening the onset time of the instability and stimulating the growth of the fluctuations to higher levels after saturation, which also confirms predictions from linear theory. 
As a result, the relaxation of the beam velocity $|v_b(\tau)/v_{Ac}|$ becomes faster and deeper, simultaneous with a strong induced (perpendicular) temperature anisotropy $A_b > 1$ (even after the saturation).
However, the (maximum) induced anisotropy $A_b >1$ is lower for denser beams, in agreement with the observational data collected by PSP in \cite{Verniero-etal-2022}, see panel (a) of Figure~2 therein.

In Figure~\ref{fig:2}, we plot, as an example, the contours of the proton velocity distribution, initially at $\tau=0$ (gray-dashed) and at the end $\tau=600$ (red-solid), corresponding to
the results from run~4 in Figure~\ref{fig:1} (second left column). 
It is evident that after the saturation of the RHPB instability, the beam population is not only relaxed to lower drift velocities but also gains a strong temperature anisotropy in the perpendicular direction, resembling those observed by PSP for the hammerhead populations \citep{Verniero-etal-2022}.

\begin{figure*}[t!]
    \centering
    \includegraphics[width=0.485\textwidth]{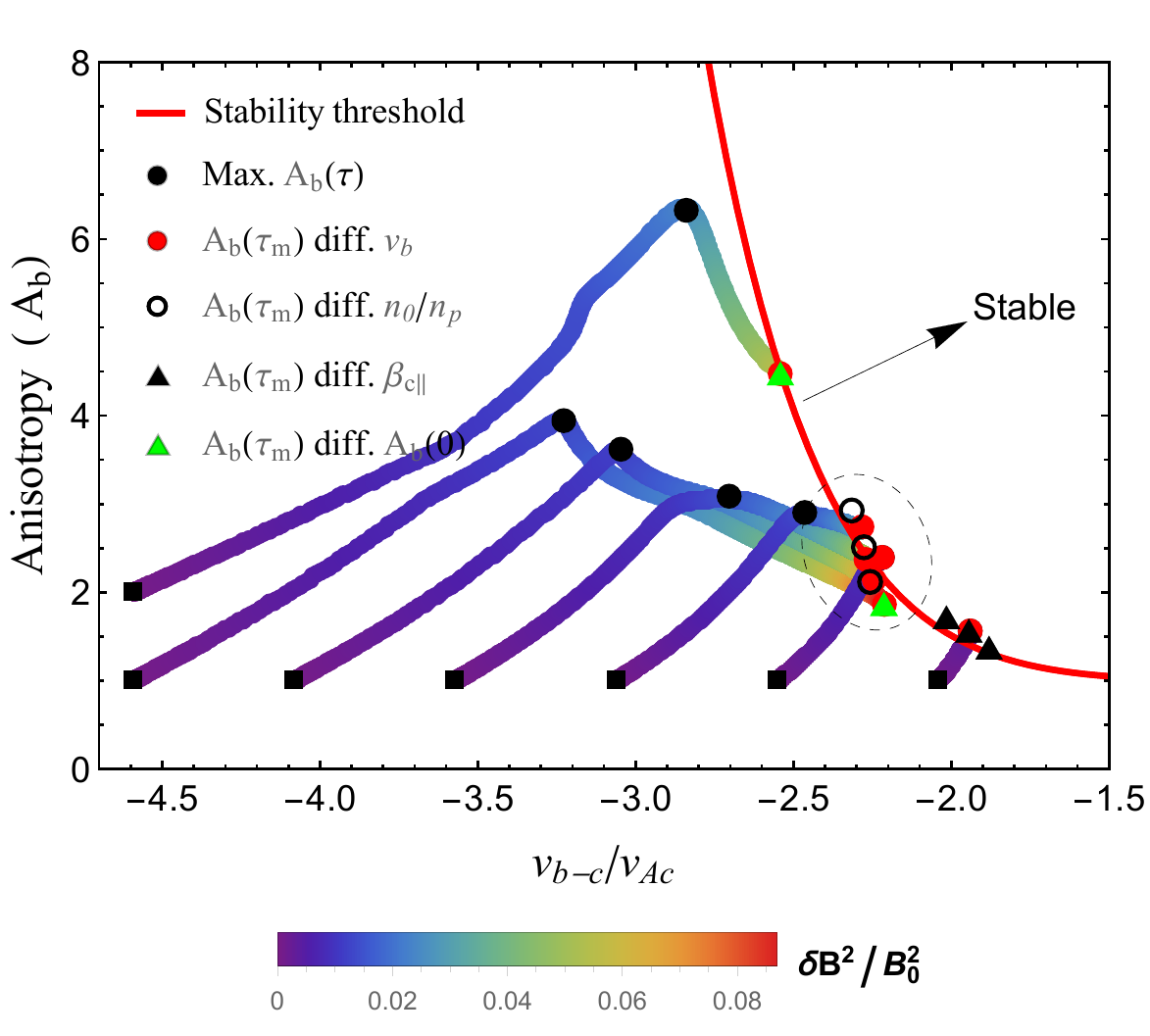}\includegraphics[width=0.51\textwidth]{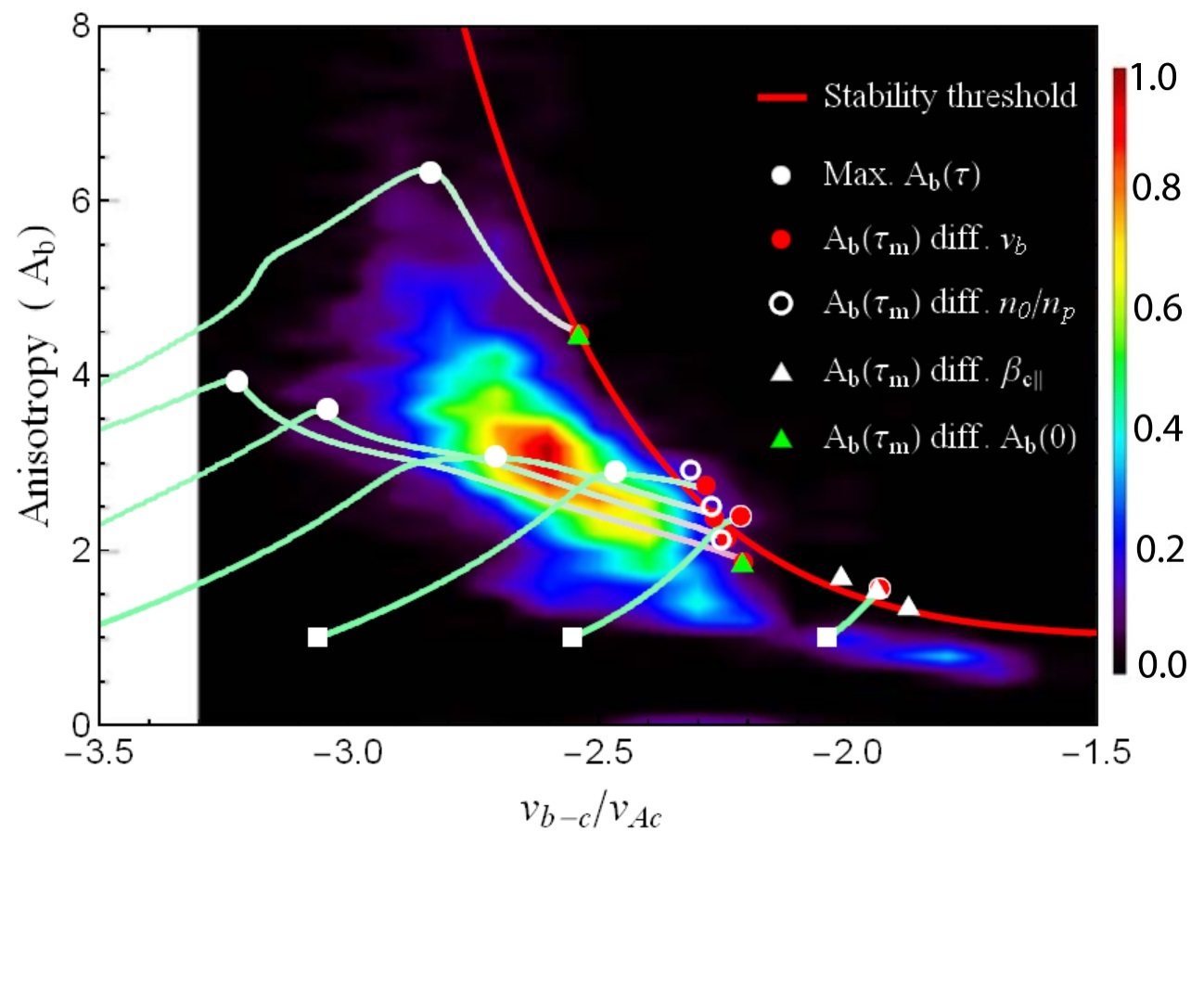}
    \caption{Left: QL dynamical paths of temperature anisotropy ($A_b$) vs. beam-core drift velocity ($v_{b-c}/v_{Ac}$). Black squares are initial states, and the magnetic wave energy level is color-coded. The final states after saturation gather on the red curve (see text): red (big) dots for runs 1-3 and 13-15, circles for 4-6, black triangles for 7-9, and green triangles for 10-12. Right: same as left panel (black symbols reversed into white symbols) but superimposed on PSP observations, with color-coded (normalized) number of events with hammerhead populations [adapted from Figure~2, panel (b), in \cite{Verniero-etal-2022} \textcopyright AAS].}
    \label{fig:3}
\end{figure*}
%

%
%
The third-left column in Figure~\ref{fig:1} displays the QL results for runs 7, 8, and 9,  for different initial values of the core parallel beta $\beta_{c\parallel}(0)=1.0$ (blue), 0.41 (orange), and 0.1 (green), respectively. 
Temporal profiles of the wave energy density (second upper row) and the proton beam parameters do not change much. However, their dependency on the initial values of $\beta_{c\parallel}(0)$ is less uniform, as also found for the linear growth rates. 
For instance, the faster onset and maximum level of the enhanced fluctuation are associated with $\beta_{c\parallel}(0)=0.41$, in agreement with the peaking growth rate (top panel) from linear theory. 
However, for the same run~8, the drift velocity relaxation and the induced temperature anisotropy (orange lines) reach intermediate values compared to the other two cases. 
Moreover, the deeper relaxation of the beam drift velocity is associated with the lowest induced anisotropy (blue lines) in run~7 with $\beta_{c\parallel}(0)=1.0$, and vice versa in run~9 (green lines) with $\beta_{c\parallel}(0)=0.1$.

%
%
The right (last) column in Figure~\ref{fig:1} displays the QL results for runs 10, 11, and 12, for different initial beam anisotropies, $A_b=2.0$ (blue), 1.0 (orange), and 0.62 (green), respectively. 
Differences obtained for the saturated levels of the unstable fluctuations (second upper panel) and the relaxation of the beam velocity (third panel) are insignificant. 
However, the induced beam temperature anisotropies for each run are notable. In particular, $A_b >6$ reached at the saturation of run~10 (which, however, starts with already an $A_b(0) = 2$), but also for run~12, reaching $A_b \simeq 4$ at the saturation.  
%

\subsection{Constraints on the observed hammerhead populations}
Dynamic paths of the beam temperature anisotropy $A_b(\tau)>1$, induced by the RHPB instability, as a function of relative beam velocity $v_{b-c}(\tau)/v_{Ac}$ are displayed for all 15 runs in Figure~\ref{fig:3}, both left and right panels. 
Initial states in the $(A_b, v_{b-c}/v_{Ac})-$space are shown with black squares, and the magnetic wave energy ($\delta B^2/ B_0^2$) level is color-coded (rainbow scheme). 
The final states after saturation are indicated with red (big) dots for runs 1, 2, 3, 13, 14, and 15, circles for runs 4, 5, and 6, black triangles for runs 7, 8, and 9, and green triangles for runs 10, 11, 12. 
These temporal profiles show the relaxation of beam drift velocity towards lower values concomitant with the heating of the (initially isotropic) beam protons in the perpendicular direction ending up with a large anisotropy $A_b(\tau)>1$, at later times after saturation. 
These effects are a direct consequence of the enhanced RHPB electromagnetic fluctuations and are consistent with the observations collected by PSP; see the right panel and Figure~2 in \cite{Verniero-etal-2022}.
In the right panel, we directly compare the dynamic paths with the observations, superimposing on the PSP events relevant for the hammerhead populations reported by \cite{Verniero-etal-2022} in their Figure~2, panel (b).
The occurrence rate of the hammerhead distributions (normalized to the total number of events and color-coded on the right-side bar) is computed during the 7~hr ion-scale wave storm. It can be interpreted as a measure of perpendicular velocity-space diffusion strength.
A negative drift velocity is chosen with respect to the orientation of the magnetic field, according to the adopted instrumental convention.

Thus, all the dynamic paths follow the same trend, passing through the specific states of the hammerhead distributions, even those with the highest occurrence rate.
After the instability saturation, the final states align along the following marginal stability condition (red curve in Figure~\ref{fig:3})
\begin{equation}\label{eq7}
 A_b=1+a(-v_{b-c}/v_{Ac})^b,
\end{equation}
with $a=0.0019$ and $b=8.04$.
These states also correspond to the highest levels of wave energy density (color-coded in the left panel).
An exceptional result is that this threshold precisely bounds the events characterized by hammerhead distributions (right panel in Figure~\ref{fig:3}).

Regimes that are stable against the RHPB instability are located on the right side of the stability threshold (red line), as indicated by the black arrow. These states occur when the drift velocities are annihilated by the temperature anisotropies in the perpendicular direction, whether both are small or both tend to be large.
On the other (left) side of the stability threshold, the RHPB instability is active, and there is a wave-particle energy exchange, which is also suggested by the increase in wave energy density. 
It is also worth mentioning that the final states of ten out of the 15 runs settle down and accumulate in relatively narrow parametric intervals $A_b = [1.8, 2.9]$, and $v_{b-c}/v_{Ac} = [-2.3, -2.2]$, indicated within a dashed-line oval.
These appear to be the most likely quasi-stable states against the RHPB instability predicted by our QL theory for the plasma conditions reported by the PSP observations.
Indeed, the mean values of the beam drift velocity and the perpendicular temperature anisotropy corresponding to the events with hammerhead proton distributions are $v_{b-c}/v_{Ac} \approx 2.5$ and $A_b\approx 2.5$ \citep{Verniero-etal-2022}, which are also in good agreement with our QL results.

\section{Conclusions}
%
PSP's new in situ data from the young solar wind should provide unprecedented details of plasma dynamics at temporal resolutions superior to other missions. 
In this Letter, we have investigated the formation conditions of the so-called hammerhead population associated with core-beam distributions of protons from PSP measurements \citep{Verniero-etal-2022}.
We propose a relatively simple QL approach for the RHPB instability that provides sufficient evidence of its involvement in the generation of such a hammerhead distribution through direct action on the proton beam relaxation.
Depending on the initial conditions chosen according to the PSP observations, the energy density of the resulting waves can reach a few percent (up to around 10 \%) of the energy density of the (uniform) magnetic field.
Moreover, the growing RHPB waves convert the bulk energy of the beam, preferentially heating the proton beam and leading to a significant temperature anisotropy in the perpendicular direction, $A_b = T_{b \perp}/T_{b \parallel} > 1$ (Figure~\ref{fig:1}).
For all runs, the core protons and the electrons also gain energies in the perpendicular direction. Still, the maximum values obtained do not exceed $12\%$ of initial values, e.g., $A_{c, e}(\tau_{max})$ increases from 1.0 to 1.12.
Our results strongly suggest that the third hammerhead population is not necessarily a distinct one but intimately related to the beam. 
The model proposed here naturally results from the diffusion of beam protons induced by the instability in the velocity space in the direction perpendicular to the magnetic field  (Figure~\ref{fig:2}).

The importance of the present results is further highlighted by the exceptional agreement with the observational data (Figure~\ref{fig:3}, right panel).
The QL dynamic paths obtained for diﬀerent initial parameters pass through the PSP events are found relevant for the presence of hammerhead populations.
Moreover, the marginal stability, Equation~\eqref{eq7}, predicted by our QL approach, shapes the margins of these unstable events very well.
Thus, we can conclude that these hammerhead distributions are rather transient states still subject to relaxation mechanisms, among which kinetic (self-generated) instabilities such as the one discussed here are very likely involved.
Future analyses with the help of numerical simulations thus become very motivating to confirm this hypothesis.


%
\section*{acknowledgments}
The authors acknowledge support from the Ruhr-University Bochum, the Katholieke Universiteit Leuven, and Qatar University. These results were also obtained in the framework of the projects C14/19/089 (C1 project Internal Funds KU Leuven), G002523N (FWO-Vlaanderen), SIDC Data Exploitation (ESA Prodex), Belspo project B2/191/P1/SWiM. 
P.H.Y. acknowledges the support by NASA Awards 80NSSC19K0827, 80NSSC23K0662, the Department of Energy grant DE-SC0022963 through the NSF/DOE Partnership in Basic Plasma Science and Engineering, NSF Grant 2203321, to the University of Maryland.


\bibliography{papers}{}
\bibliographystyle{aasjournal}
\end{document}